\documentclass[twocolumn,showpacs,
aps,pra,notitlepage,longbibliography,nofootinbib]{revtex4-1}

\usepackage{xcolor}
\usepackage{graphicx}
\usepackage{amsmath}
\usepackage{amsfonts}
\usepackage{amssymb}
\usepackage{maybemath}
\usepackage{bbm}
\usepackage{hyperref}

\usepackage{maybemath}

\usepackage{bbm}

\usepackage[utf8]{inputenc}
\usepackage[cal=dutchcal,scr=dutchcal]{mathalfa}

\def\calL{{\mathcal L}}

\def\calh{{\mathcal{h}}}

\def\dd{{\mathrm d}}
\def\ii{{\mathrm i}}
\def\ee{{\mathrm e}}

\def\gg{{\mathrm{g}}}

\definecolor{garrosgreen}{rgb}{0.1, 0.4, 0.1}
\definecolor{dartmouthgreen}{rgb}{0.05, 0.5, 0.06}
\definecolor{duelferred}{rgb}{0.7, 0.2, 0.1}
\definecolor{cambridgeblue}{rgb}{0.1, 0.3, 1.0}
\definecolor{oxfordblue}{rgb}{0.05, 0.2, 0.7}

\newcommand{\vapprox}{\; \dot{\approx} \;}

\begin{document}

%
%
\title{Fifth Force and Hyperfine Splitting in Bound Systems}

\author{Ulrich D. Jentschura}

\affiliation{\addrRolla}

\newcommand{\addrRolla}{Department of Physics, Missouri University of Science
and Technology, Rolla, Missouri 65409, USA}

\begin{abstract}
Two recent experimental observations
at the ATOMKI Institute of the Hungarian Academy of
Sciences (regarding the angular emission pattern of 
electron-positron pairs from 
nuclear transitions from excited 
states in ${}^8{\rm Be}$ and ${}^4{\rm He}$)
indicate the possible existence of 
a particle of a rest mass energy 
of roughly $17\,{\rm MeV}$. The so-called 
X17 particle constitutes 
a virtual state in the process, preceding the 
emission of the electron-positron pair.
Based on the symmetry of the nuclear transitions
($1^+ \to 0^+$ and $0^- \to 0^+$),
the X17 could either be 
a vector, or a pseudoscalar particle.
Here, we calculate the effective potentials 
generated by the X17,
for hyperfine interactions in simple atomic 
systems, for both the pseudoscalar as well as
the vector X17 hypotheses. 
The effective Hamiltonians are obtained in a general form which is
applicable to both electronic as well as muonic
bound systems.
The effect of virtual annihilation and its contribution
to the hyperfine splitting also is considered.
Because of the short range of the X17-generated 
potentials, the most promising 
pathway for the observation of the X17-mediated 
effects in bound systems concerns hyperfine interactions,
which, for $S$ states, are given by modifications
of short-range (Dirac-$\delta$) potentials
in coordinate space. For the pseudoscalar 
hypothesis, the exchange of one virtual X17 quantum
between the bound lepton and the nucleus 
exclusively leads to hyperfine effects, but 
does not affect the Lamb shift.
Effects due to the X17 are shown to be drastically 
enhanced for muonic bound systems.
Prospects for the detection of 
hyperfine effects mediated by X17 exchange are 
analyzed for muonic deuterium, muonic hydrogen,
muonium, true muonium ($\mu^+\mu^-$ bound system),
and positronium.
\end{abstract}

\maketitle

%
%
\section{Introduction}
\label{sec1}

For decades, atomic physicists have tried to 
push the accuracy of experiments and 
theoretical predictions of transitions in 
simple atomic systems higher~\cite{Ha2006}. 
The accurate measurements have led to 
stringent limits on the time variation of 
fundamental constants~\cite{FiEtAl2004,GoEtAl2014,HuEtAl2014},
and enabled us to determine 
a number of important fundamental physical
constants~\cite{MoNeTa2016}
with unprecedented accuracy.
Yet, a third motivation 
(see, e.g., Refs.~\cite{TSYa2011,BaEtAl2012}), hitherto not crowned with 
success, has been the quest to find signs of a possible 
low-energy extension of the Standard Model,
based on a deviation of experimental results
and theoretical predictions.

Recently, the possible existence of a fifth-force particle, 
commonly referred to as the ``X17'' particle because of the 
observed rest mass of $16.7\,{\rm MeV}$, 
has been investigated in Refs.~\cite{KrEtAl2016,KrEtAl2017woc2,KrEtAl2019},
based on a peak in the emission spectrum 
of electron-positron pairs in nuclear transitions
of excited helium and beryllium nuclei.
Two conceivable theoretical explanations have been put forward,
both being based on low-energy additions to the 
Standard Model. The first of these involves
a vector particle (a ``massive, dark photon'',
see Refs.~\cite{FeEtAl2016,FeEtAl2017}),
and the second offers
a pseudoscalar particle (see Ref.~\cite{ElMo2016}),
which couples to light fermions as well as hadrons.

The findings of Refs.~\cite{KrEtAl2016,KrEtAl2017woc2,KrEtAl2019}
have not yet been confirmed by any other 
experiment and remain to be independently 
verified (for an overview of 
other experimental searches and  
conceivable alternative interpretations of the 
ATOMKI results, see Refs.~\cite{BaEtAl2018,Ko2020}).
However, we believe that, with the advent of 
consistent observations in two nuclear transitions
in ${}^8{\rm Be}$ and ${}^4{\rm He}$,
it is justified to carry out a calculation
of the effects induced by the X17 boson in atomic 
systems. In more general terms, we ask the 
question which effects could be expected 
from a potential ``light pseudoscalar Higgs''-type
particle in atomic spectra, as envisioned in Ref.~\cite{ElMo2016}.

Somewhat unfortunately, the rest mass range of $16.7\,{\rm MeV}$
makes the X17 particle hard to detect in 
atomic physics experiments.
The observed X17 rest mass energy is larger than
the binding energy scale for both
electronic as well as muonic bound systems~\cite{JeNa2018}.
Even more importantly, the Compton wavelength of the X17 particle 
(about $11.8\,{\rm fm}$) is 
smaller than the effective Bohr radius
for both electronic as well as muonic bound systems.
Because the Compton wavelength of the X17 particle
determines the range of the Yukawa potential,
the effects of the X17 are hard to distinguish
from nuclear-size effects 
in atomic spectroscopy experiments~\cite{JeNa2018}.

We recall that the Bohr radius amounts to
$a_0 \sim 5 \times 10^4 \, {\rm fm}$, while the
effective Bohr radius of a 
muonic hydrogen atom is 
$a_0 \sim \hbar/(\alpha m_\mu c) \sim 256 \, {\rm fm}$.
It is thus hard to find an atomic system,
even a muonic one, where one could hope to 
distinguish the effect of the X17 particle
on the Lamb shift
from the nuclear-finite-size correction
to the energy.
A possible circumvention has been discussed in
Ref.~\cite{JeNa2018}, based on a muonic carbon ion,
where the effective Bohr radius approaches the 
range of the Yukawa potential induced by the X17,
in view of the larger nuclear charge number.
However, it was concluded in Ref.~\cite{JeNa2018}
that considerable additional effort 
would be required in terms of an accurate
understanding of nuclear-size effects,
before the X17 signal could be extracted reliably.

The definition of the Lamb shift $\calL$, as envisaged in
Ref.~\cite{SaYe1990} and used in many other 
places, e.g., in Eq.~(67) of Ref.~\cite{JePa1996},
explicitly excludes hyperfine effects.
Conversely, hyperfine effects, at least 
for $S$ states, are induced, in leading order,
by the Dirac-$\delta$
peak of the magnetic dipole field of the atomic nucleus
at the origin [see Eq.~(9) of Ref.~\cite{JeYe2006}].
The Fermi contact interaction, which
gives rise to the leading-order contribution
to the hyperfine splitting for $S$ states, is proportional
to a Dirac--$\delta$ in coordinate space,
commensurate with the fact that the atomic
nucleus has a radius not exceeding the femtometer scale.
The effect of short-range potentials is thus less suppressed
when we consider the hyperfine splitting,
as compared to the Lamb shift.
The {\em afici\'{o}nados} of bound states thus 
realize that, if we consider hyperfine effects,
we have a much better chance of extracting the
effect induced by the X17, which, on the ranking scale
of the contributions, occupies a much higher 
place than for the Lamb shift alone.
Despite the large mass $m_X$ of the X17 particle,
which leads to a short-range potential 
proportional to $\exp(-m_X \, r)$ 
(in natural units with $\hbar = c = \epsilon_0 =1$,
which will be used throughout the current paper),
the effect of the X17 could thus be visible in the 
hyperfine splitting in muonic atoms.

Here, we shall elaborate on this idea,
and derive the leading corrections to the 
hyperfine splitting of $nS$, $nP_{1/2}$ 
and $nP_{3/2}$ states in ordinary as well 
as muonic hydrogenlike systems,
due to the X17 particle,
by matching the nuclear-spin dependent terms
in the scattering amplitude with the 
effective Hamiltonian.
Anticipating some results, we can say that the 
relative correction (expressed in terms of the leading Fermi 
term) is proportional to $m_r/m_X$, where $m_r$ is the 
reduced mass of the two-body bound system, while $m_X$ 
is the X17 boson mass. 
The effect is thus enhanced for muonic in comparison to electronic
bound systems.

This paper is organized as follows.
In Sec.~\ref{sec2}, we summarize the 
interaction Lagrangians for both
a hypothetical X17 vector exchange~\cite{FeEtAl2016,FeEtAl2017},
as well as a pseudoscalar exchange~\cite{ElMo2016},
with corresponding conventions for the coupling 
parameters. In Sec.~\ref{sec3}, we derive the effective 
hyperfine Hamiltonians for both  
vector and pseudoscalar exchanges.
In Sec.~\ref{sec4}, we evaluate general expressions
for the corrections to hyperfine energies induced 
by the X17 particle, for $S$ and $P$ states.
In Sec.~\ref{sec5}, we derive bounds on the 
coupling parameters for both models
in the muon sector, based on the muon $g$ factor.
Finally, in Sec.~\ref{sec6},
we apply the obtained results to muonic hydrogen,
muonic deuterium, muonium, true muonium (bound $\mu^+\mu^-$ system),
and positronium.
We also discuss the measurability of the X17 effects 
in the hyperfine structure of the mentioned 
atomic systems.
Conclusions are reserved for Sec.~\ref{sec7}.

%
%
\section{Interaction Lagrangians}
\label{sec2}

In the following, we intend to study both the 
interaction of X17 vector and pseudoscalar particles
with bound leptons (electrons and muons)
and nucleons (protons and deuterons).
Vector interactions will be denoted by the 
subscript $V$, while pseudoscalar interactions
will carry the subscript $A$, as is customary in the 
particle physics literature.
We write the interaction Lagrangian $\calL_{X,V}$
for the interaction of 
an X17 vector boson with the fermion fields
$f = e, \mu$ (electron and muon) 
and the nucleons $N = p,n$ (proton and neutron) as follows,
\begin{subequations}
\label{LXV}
\begin{equation}
\calL_{X,V} = 
-\sum_f \varepsilon_f e \, {\bar \psi}_f \, \gamma^\mu \, \psi_f  X_\mu 
-\sum_N \varepsilon_N e \, {\bar \psi}_N \, \gamma^\mu \, \psi_N X_\mu \,,
\end{equation}
where we follow the conventions delineated in the remarks
following Eq.~(1) of Ref.~\cite{FeEtAl2016} and 
Eq.~(10) of Ref.~\cite{FeEtAl2017}.
Here, $\varepsilon_f$ and $\varepsilon_N$ are the 
flavor-dependent coupling parameters for the fermions 
and nucleons, while $e =-\sqrt{4 \pi \alpha} = -0.091$ is the electron charge. 
The fermion and nucleon field-operators
(the latter, interpreted as field operators for the 
composite particles) and denoted as 
$\psi_f$ and $\psi_N$, while the $X_\mu$ is the X17 field operator.
For reasons which will become obvious later, we
use, in Eq.~\eqref{LXV}, the alternative conventions,
\begin{equation}
\calh'_f = \varepsilon_f \, e \,,
\qquad
\calh'_N = \varepsilon_N \, e \,,
\end{equation}
\end{subequations}
for the coupling parameters to the hypothetical
X17 vector boson.
Our conventions imply that for $\varepsilon_N > 0$, the
coupling parameter $\calh'_N$ parameterizes
a ``negatively charged'' nucleon under the additional
$U(1)$ gauge group of the vector $X$ particle.

According to a remark following the text after Eq.~(9) of 
Ref.~\cite{FeEtAl2017},
conservation of $X$ charge implies that the couplings to the proton and
neutron currents fulfill the relationships
\begin{equation}
\label{eps_proton_neutron}
\varepsilon_p = 2 \varepsilon_u + \varepsilon_d \,,
\qquad
\varepsilon_n = \varepsilon_u + 2 \varepsilon_d \,,
\end{equation}
where the up and down quark couplings are denoted by 
the subscripts $u$ and $d$.
Numerically, one finds [see the detailed discussion
around Eqs.~(38) and~(39) of Ref.~\cite{FeEtAl2017}]
that the electron-positron field coupling $\varepsilon_e$ needs
to fulfill the relationship
\begin{equation}
\label{epsEbound}
2 \times 10^{-4} < \varepsilon_e < 1.4 \times 10^{-3} \,.
\end{equation}
Furthermore, in order to explain the experimental
observations~\cite{KrEtAl2016,KrEtAl2017woc2},
one needs the neutron coupling to fulfill
[see Eq.~(10) of Ref.~\cite{FeEtAl2016}]
\begin{equation} 
\label{valuen}
|\varepsilon_n| = |\varepsilon_u + 2 \varepsilon_d |
\approx \left| \frac32 \, \varepsilon_d \right| 
\approx \frac{1}{100} \,.
\end{equation}
Because the hypothetical $X$ vector particle acts like a 
``dark photon'' which is hardly distinguishable
from the ordinary photon in the high-energy domain,
the proton coupling $\varepsilon_p$ is highly constrained.
According to Eq.~(8) and~(9) of Ref.~\cite{FeEtAl2017}, 
and Eq.~(35) of Ref.~\cite{FeEtAl2017}, one needs to have
\begin{equation} 
\label{constrainp}
|\varepsilon_p| = |2 \varepsilon_u + \varepsilon_d | \lesssim 
8 \times 10^{-4} \,.
\end{equation} 
This is why the conjectured X17 vector boson 
is referred to as ``protophobic''
in Refs.~\cite{FeEtAl2016,FeEtAl2017}.

Following Ref.~\cite{ElMo2016}, we write the 
interaction Lagrangian for the fermions interacting with 
the pseudoscalar candidate of the X17 particle as follows,
\begin{equation} 
\label{LXA}
\calL_{X,A} = 
-\sum_f \calh_f \; {\bar \psi}_f \, \ii \, \gamma^5 \, \psi_f \, A
-\sum_N \calh_N \; {\bar \psi}_N \, \ii \, \gamma^5 \, \psi_N \, A \,,
\end{equation} 
where $A$ is the field operator of the pseudoscalar field.
Inspired by an analogy with putative 
pseudoscalar Higgs couplings~\cite{ChCh2012},
the pseudoscalar couplings have been estimated in 
Refs.~\cite{ChCh2012,ElMo2016} to be of the 
functional form
\begin{equation}
\calh_f = \xi_f \, \frac{m_f}{v} \,,
\qquad
\calh_N = \xi_N \, \frac{m_N}{v} \,,
\end{equation}
where $v = 246 \, {\rm GeV}$ is the vacuum expectation value of the 
Higgs (or Englert--Brout--Higgs, see Refs.~\cite{EnBr1964,Hi1964})
field, $m_f$ is the fermion mass, and $m_N$ is the nucleon's mass.
Furthermore, the parameters 
$\xi_f$ and $\xi_N$ could in principle be assumed
to be of order unity.
Note that the spin-parity of the Standard Model Higgs boson
has recently been determined to be consistent with a 
scalar, not pseudoscalar, particle~\cite{AaEtAl2015},
but it is still intuitively suggested to parameterize the 
couplings to the novel putative pseudoscalar X17 in the 
same way as one would otherwise parameterize the 
couplings to the Higgs particle.

According to Eq.~(2.7) and the remark following 
Eq.~(3.12) of Ref.~\cite{ElMo2016}, the nucleon 
couplings can roughly be estimated as
\begin{subequations}
\label{hpn}
\begin{align}
\label{hp}
\calh_p =& \; \frac{m_p}{v} \left( -0.40 \, \xi_u - 1.71 \, \xi_d \right) 
\approx -2.4 \times 10^{-3} \,,
\\[0.1133ex]
\label{hn}
\calh_n =& \; \frac{m_n}{v} \left( -0.40 \, \xi_u + 0.85 \, \xi_d \right) 
\approx 5.1 \times 10^{-4} \,,
\end{align}
\end{subequations}
where we have assumed $\xi_u \approx \xi_d \approx 0.3$.
For the electron-positron field, 
based on other constraints detailed in Ref.~\cite{ElMo2016}, one has 
to require that [see Eq.~(4.2) of Ref.~\cite{ElMo2016}]
\begin{equation}
\label{xi_e_lower}
\xi_e \mathop{>}^{\mbox{!}} 4 \,,
\qquad
\calh_e \mathop{>}^{\mbox{!}} \frac{4 \, m_e}{v} = 8.13 \times 10^{-6} \,.
\end{equation}
Based on a combination of experimental data~\cite{AnEtAl2015}
and theoretical considerations~\cite{AlWe2018,LiWaWa2019,ElPriv2020}, one
can also derive an upper bound,
\begin{equation}
\label{xi_e_upper}
\xi_e \mathop{<}^{\mbox{!}} 500 \,,
\qquad
\calh_e \mathop{<}^{\mbox{!}} \frac{500 \, m_e}{v} = 10^{-3} \,,
\end{equation}
which will be used in the following.

\begin{figure}[t!]
\begin{center}
\begin{minipage}{0.91\linewidth}
\begin{center}
\includegraphics[width=1.0\linewidth]{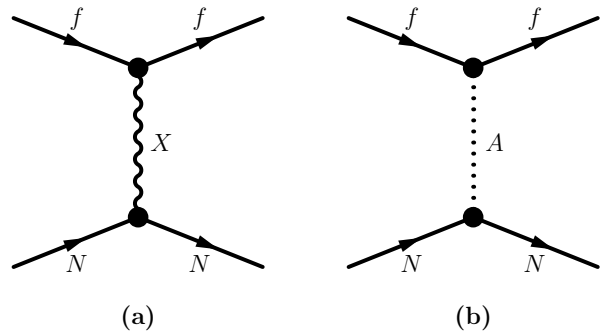}
\caption{\label{fig1} The one-quantum exchange 
scattering amplitude for the X17 particle is matched 
against the effective Hamiltonian,
for the vector hypothesis [diagram (a)] and the 
pseudoscalar hypothesis [diagram (b)].
The arrow of time is from left to right.}
\end{center}
\end{minipage}
\end{center}
\end{figure}

%
%
\section{Matching of the Scattering Amplitude}
\label{sec3}

In order to match the scattering amplitude 
(see Fig.~\ref{fig1})
with the effective Hamiltonian,
we use the approach outlined in Chap.~83 of 
Ref.~\cite{BeLiPi1982vol4},
but with a slightly altered normalization
for the propagators, better adapted to 
natural unit system ($\hbar = c = \epsilon_0 = 1$).
Specifically, we use the bispinors in the 
representation [cf.~Eq.~(83.7) of Ref.~\cite{BeLiPi1982vol4}]
\begin{equation}
\label{defu}
u_{f,N} = \left( \begin{array}{c} 
\left( 1 - \dfrac{\vec p_{f,N}^{\,2}}{8 m^2} \right) w_{f,N} \\[3ex]
\dfrac{ \vec \sigma \cdot \vec p}{2 m} \, w_{f,N} 
\end{array} \right) \,,
\end{equation}
where $f,N$ stands for the bound fermion,
or the nucleus, and $w_{f,N}$ are the nonrelativistic 
spinors. Of course, 
two two-component {\em (sic!)} spinors
constitute the four-component bispinor $u_{f,N}$
of the same field.
The massive photon propagator (for the X17 vector 
hypothesis) is used in the following normalization
(we may ignore the frequency of the photon
in the order of approximation
relevant for the current article),
\begin{subequations}
\begin{align}
\label{D00}
D_{00}(\vec q) =& \; - \frac{1}{\vec{q}^{\,2} + m_X^2} \,,
\\[0.1133ex]
\label{Dij}
D_{ij}(\vec{q}) =& \; - \frac{1}{\vec{q}^{\,2} + m_X^2}\,
\left[ \delta^{ij} - \frac{q^i \, q^j}{\vec{q}^{\,2} + m_X^2} \right] \,.
\end{align}
\end{subequations}
The derivation of the massive vector boson propagator 
in the Coulomb gauge, which is best adapted to bound-state 
calculations and
involves a certain subtlety, is discussed 
in Appendix~\ref{appa}.
The pseudoscalar propagator is used in the normalization
\begin{equation}
D_A(\vec q) = - \frac{1}{\vec{q}^{\,2} + m_X^2} \,,
\end{equation}
where we also ignore the frequency.
The scattering amplitude for the X17 vector particle reads as
\begin{align}
M_{fi,V} =& \; \calh_f \, \calh_N \,
\left\{ 
\left( \bar u'_f \, \gamma^0 \, \bar u_f \right) \,
\left( \bar u'_N \, \gamma^0 \, \bar u_N \right) \, D_{00}
\right.
\nonumber\\[0.1133ex]
& \; +
\left.
\left( \bar u'_f \, \gamma^i \, \bar u_f \right) \,
\left( \bar u'_N \, \gamma^j \, \bar u_N \right) \, D_{ij}
\right\} \,,
\end{align}
and
\begin{equation}
M_{fi,A} = \calh_f \, \calh_N \,
\left( \bar u'_f \, \ii \, \gamma^5 \, \bar u_f \right) \,
\left( \bar u'_N \, \ii \, \gamma^5 \, \bar u_N \right) \, D_A \,,
\end{equation}
for the pseudoscalar case.
Here, we denote the final states of the 
scattering process by a prime, $u'_f = u_f(\vec p'_f)$, $u'_N = u_N(\vec p'_N)$,
while the initial states are
$u_f = u_f(\vec p_f)$ and $u_N = u_N(\vec p_N)$,
and the bar denotes the Dirac adjoint.
Analogous definitions are used for the 
$w'_{f,N}$ and $w_{f,N}$ in Eq.~\eqref{defu}.
Furthermore, we have $\vec p_f + \vec p_N = {\vec p\,}'_f + {\vec p\,}'_N$.
The momentum transfer is $\vec q = {\vec p\,}'_f - \vec p_f =
\vec p_N - {\vec p\,}'_N$.

The form~\eqref{defu} is valid for the bispinors
if the Dirac equation is solved in the Dirac representation
of the Dirac matrices,
\begin{align}
\gamma^0 =& \; \left( \begin{array}{cc} \mathbbm{1}_{2 \times 2} & 0 \\
0 & -\mathbbm{1}_{2 \times 2} \end{array} \right) \,,
\quad
\gamma^i = \left( \begin{array}{cc} 0 & \sigma^i \\
-\sigma^i & 0 \end{array} \right) \,,
\nonumber\\[0.1133ex]
\gamma^5 =& \; \left( \begin{array}{cc} 0 & \mathbbm{1}_{2 \times 2} \\
\mathbbm{1}_{2 \times 2} & 0 \end{array} \right) \,.
\end{align}
The scattering amplitudes
are matched against the effective Hamiltonian by the relation
\begin{equation}
M_{fi} = - (w'^+_{f} \, w'^+_{N}) \,
U(\vec p_f, \vec p_N, \vec q) \,  
(w_{f} \, w_{N}) \, ,
\end{equation}
where $U(\vec p_f, \vec p_N, \vec q)$ is the the effective Hamiltonian.
The scattering amplitude $M_{fi}$ is a matrix element
involving four spinors, two of which represent
the final and initial states of the two-particle system.

The scattering amplitude, 
evaluated between four spinors 
[cf.~Eq.~(83.8) of Ref.~\cite{BeLiPi1982vol4}],
must now be matched against a Hamiltonian which acts on only
one wave function in the end.
We need to remember that the scattering amplitude
corresponds to a matrix element of the
Hamiltonian. 
Any matrix element of the Hamiltonian, even in the one-particle setting, is
sandwiched between two wave functions, not one.
Then, going into the center-of-mass frame
$\vec q = {\vec p\,}'_f - \vec p_f = \vec p_N - {\vec p\,}'_N$
means that the wave function is written
in terms of a center-of-mass coordinate
$\vec R$, and a relative coordinate $\vec r$.
In the center-of-mass frame, one eliminates
the dependence on the center-of-mass coordinate
$\vec R$ and the total momentum $\vec P = \vec p_f + \vec p_N$.
Fourier transformation under the condition
${\vec p\,}'_f + {\vec p\,}'_N = \vec p_f + \vec p_N$
leads to the effective Hamiltonian 
[cf.~Eq.~(83.15) of Ref.~\cite{BeLiPi1982vol4}].

For the record, we note that in the 
X17 vector case, the $00$ component of the 
photon propagator gives the leading, spin-independent
term in the effective Hamiltonian,
\begin{equation}
H_0 = \frac{\calh'_f \, \calh'_N}{4 \pi r} \, \exp(-m_X \, r) \,.
\end{equation}
Under the replacements $\calh'_f \to e$ and $\calh'_N \to -e$,
in the massless limit $m_X \to 0$, 
one recovers the Coulomb potential,
$H_0 \to -\frac{e^2}{4 \pi r} = -\frac{\alpha}{r}$.
One finally extracts the terms responsible for the hyperfine
structure, i.e., those involving the nuclear 
spin operator $\vec\sigma_N$, and obtains the 
following hyperfine Hamiltonian for a vector X17 particle,
\begin{multline}
\label{HHFS1}
H_{{\rm HFS},V} = 
\frac{\calh'_f \, \calh'_N}{16 \, \pi \, m_f \, m_N} \,
\left[ -\frac{8 \pi}{3} \delta^{(3)}(\vec r) \, 
\vec\sigma_f \cdot \vec\sigma_N
\right.
\\[0.1133ex]
- \frac{m_X^2 \, 
\left( \vec\sigma_f \cdot \vec r \; \vec\sigma_N \cdot \vec r - 
r^2 \, \vec\sigma_f \cdot \vec\sigma_N \right)}{r^3} \, \ee^{-m_X \, r}
\\[0.1133ex]
- \left( 1 + m_X \, r \right)
\frac{ 3 \, \vec\sigma_f \cdot \vec r \; \vec\sigma_N \cdot \vec r - 
r^2 \, \vec\sigma_f \cdot \vec\sigma_N }{r^5} \, \ee^{-m_X \, r}
\\[0.1133ex]
\left.
- \left( 2 + \frac{m_f}{m_N} \right) \,
\left( 1 + m_X \, r \right) \,
\frac{\vec\sigma_N \cdot \vec L}{r^3} 
 \ee^{-m_X \, r} \right] \,.
\end{multline}
Taking the limit $m_X \to 0$, and
replacing
\begin{equation}
\calh'_f \to e \,,
\qquad
\calh'_N \to \frac{\gg_N \,(-e)}{2} =
\frac{\gg_N \,|e|}{2} \,,
\qquad
e^2 = 4 \pi \alpha \,,
\end{equation}
one recovers the Fermi Hamiltonian $H_F$ 
[see Eq.~(10) of Ref.~\cite{JeYe2006}],
\begin{align}
\label{HSLEAD}
H_F =& \; \frac{\gg_N \, \alpha}{m_f \, m_N} \, 
\left[ \frac{\pi}{3} \, \vec\sigma_f \cdot \vec\sigma_N \,
\delta^{(3)}(\vec r) 
\right.
\nonumber\\[0.1133ex]
& \; \left. + \frac{ 3 \, \vec\sigma_f \cdot \vec r \; \vec\sigma_N \cdot \vec r -
r^2 \, \vec\sigma_f \cdot \vec\sigma_N }{8 \, r^5} 
+ \frac{\vec\sigma_N \cdot \vec L}{4 \, r^3} 
\right] \,,
\end{align}
where we have ignored the reduced-mass correction
proportional to $m_f/m_N$ in the $\vec\sigma_N \cdot \vec L$ 
term in Eq.~\eqref{HHFS1}.
For a pseudoscalar exchange, one has
\begin{multline}
\label{HHFS2}
H_{{\rm HFS},A} =
\frac{\calh_f \, \calh_N}{16 \, \pi \, m_f \, m_N} \,
\left[ 
\frac{4 \pi}{3} \delta^{(3)}(\vec r) \, 
\vec\sigma_f \cdot \vec\sigma_N
\right.
\\[0.1133ex]
- \frac{m_X^2 \, \vec\sigma_f \cdot \vec r \;
\vec\sigma_N \cdot \vec r}{r^3} \, \ee^{-m_X \, r}
\\[0.1133ex]
\left.
+ \left( 1 + m_X \, r \right)
\frac{3 \, \vec\sigma_f \cdot \vec r \, \vec\sigma_N \cdot \vec r
- \vec\sigma_f \cdot \vec\sigma_N \, r^2}{r^5} \, \ee^{-m_X \, r}
\right] \,.
\end{multline}
Note that the Hamiltonian
given in Eq.~\eqref{HHFS2} constitutes the complete
Hamiltonian derived from pseudoscalar exchange,
which, in view of the $\gamma^5$ matrix in the Lagrangian
given in Eq.~\eqref{LXA}, contributes only to the 
hyperfine splitting, but not to the Lamb shift,
in leading order [i.e., via the exchange of 
one virtual particle, as given in Fig.~\ref{fig1}(a)].
For a deuteron nucleus, 
the spin matrix $\vec\sigma_N$ has to be replaced by
$2 \, \vec I_N$, where $\vec I_N$ is the 
spin operator of the deuteron, corresponding to the 
spin-1 particle.
Important bounds on the coupling parameters 
$\calh'_\mu$ and $\calh_\mu$ can be
derived from the muon anomalous magnetic moment (see Fig.~\ref{fig2}).

%
%
\section{Hyperfine Structure Corrections}
\label{sec4}

%
%
\subsection{X17 Boson Exchange}

In order to analyze the $S$ state hyperfine splitting,
we extract from Eqs.~\eqref{HHFS1} and~\eqref{HHFS2}
the terms which are nonzero when evaluated on a 
spherically symmetric wave function.
This entails the replacements
\begin{subequations}
\begin{align}
\vec\sigma_f \cdot \vec r \; 
\vec\sigma_N \cdot \vec r 
\to & \; \frac13 \, r^2 \, \vec\sigma_f \cdot \vec\sigma_N \,,
\qquad
\vec\sigma_N \cdot \vec L \to  0 \,,
\\[0.1133ex]
\label{HHFSV}
H_{{\rm HFS},V} \to & \;
-\frac{\calh'_f \, \calh'_N \,
\vec\sigma_f \cdot \vec\sigma_N}{24\, \pi \, m_f \, m_N} 
\nonumber\\[0.1133ex]
& \; \times \left[
4 \pi \, \delta^{(3)}(\vec r) \,
- \frac{m_X^2}{r} \, \ee^{-m_X \, r}
\right] \,,
\\[0.1133ex]
\label{HHFSA}
H_{{\rm HFS},A} \to & \;
\frac{\calh_f \, \calh_N \,
\vec\sigma_f \cdot \vec\sigma_N}{48 \, \pi \, m_f \, m_N} 
\nonumber\\[0.1133ex]
& \; \times \left[
4 \pi \, \delta^{(3)}(\vec r) \,
- \frac{m_X^2}{r} \, \ee^{-m_X \, r}
\right] \,.
\end{align}
\end{subequations}
The expectation value of the Fermi Hamiltonian is 
\begin{equation}
\label{EFERMI}
E_F(nS) = 
\left< nS_{1/2} | H_F | nS_{1/2} \right> = 
\gg_N \, \frac{\alpha \, (Z\alpha)^3 \, m_r^3}{3 \, n^3 \, m_f \, m_N} \,
\left< \vec\sigma_f \cdot \vec\sigma_N \right> \,.
\end{equation}
Here, $Z$ is the nuclear charge number,
and $m_r = m_f m_N/(m_f + m_N)$ is the reduced mass 
of the system. 
We use the nuclear $g$ factor in the normalization 
\begin{equation}
\label{defg}
\vec\mu_N = \gg_N \, \frac{|e|}{2 m_N} \, \frac{\vec\sigma_N}{2} \,,
\end{equation}
which can more easily be extended to more general two-body 
systems than a definition in terms on the nuclear magneton.
For the proton, one has $\gg_p = 5.5856\dots$ as the proton's $g$ 
factor~\cite{DSGa2012,MoEtAl2014},
while the definition~\eqref{defg} implies that 
$\gg_d = 1.713\dots$ for the deuteron~\cite{MoNeTa2016}.
For true muonium ($\mu^+\mu^-$ bound system) and 
positronium, one has $\gg_N = 2$ according to the 
definition~\eqref{defg}.

By contrast, in the limit $m_X \to \infty$, one verifies that 
\begin{equation}
\lim_{m_X \to \infty} 
\left\{
\frac{m_X^2}{r} \, \ee^{-m_X \, r} 
\right\}
= 4 \pi \, \delta^{(3)}(\vec r) \,,
\end{equation}
and the two Hamiltonians given in Eqs.~\eqref{HHFSV}
and~\eqref{HHFSA} vanish in the limit of an infinitely heavy X17 particle.
This implies that the expectation values of $S$ states
of the Hamiltonians in Eqs.~\eqref{HHFSV} and~\eqref{HHFSA}
have to carry at least one power of $m_X$ in the denominator,
and in particular, that the correction to the 
hyperfine energy will be of order $\alpha (Z\alpha)^4$,
not $\alpha (Z\alpha)^3$, as one would otherwise 
expect from the two individual terms in 
Eqs.~\eqref{HHFSV} and~\eqref{HHFSA}.
For the vector hypothesis,
one finds that $E_{X,V}(nS_{1/2}) = 
\left< nS_{1/2} | H_{{\rm HFS},V} | nS_{1/2} \right>$,
for the leading and subleading terms
in the expansion in inverse powers of $m_X$,
can be expressed as
\begin{multline}
\label{EXVnS12}
E_{X,V}(nS_{1/2}) = 
\calh'_f \, \calh'_N \, 
\left( 
- \frac{2 (Z\alpha)^4}{3 \pi n^3} \, \frac{m_r^4}{m_f \, m_N \, m_X} 
\right.
\\
\left.
+ \frac{5 (Z\alpha)^5}{3 \pi n^3} \, 
\left( 1 + \frac{1}{5 n^2} \right) \,
\frac{m_r^5}{m_f \, m_N \, m_X^2} 
\right)
\left< \vec\sigma_f \cdot \vec\sigma_N \right>_{S_{1/2},F} \,.
\end{multline}
We have neglected relative corrections
of higher than first order in $\alpha \, m_f/m_X$ and $m_f/m_N$.
For the pseudovector hypothesis, one finds
that $E_{X,A}(nS_{1/2}) = 
\left< nS_{1/2} | H_{{\rm HFS},A} | nS_{1/2} \right>$
is given as follows,
\begin{multline}
\label{EXAnS12}
E_{X,A}(nS_{1/2}) = 
\calh_f \, \calh_N \, 
\left( 
\frac{(Z \alpha)^4}{3 \pi n^3} \, \frac{m_r^4}{m_f \, m_N \, m_X} 
\right.
\\
\left.
- \frac{5 (Z \alpha)^5}{6 \pi n^3} \, 
\left( 1 + \frac{1}{5 n^2} \right) \,
\frac{m_r^5}{m_f \, m_N \, m_X^2} 
\right)
\left< \vec\sigma_f \cdot \vec\sigma_N \right>_{S_{1/2},F} \,.
\end{multline}
The $S$ state splitting is obtained from 
the following expectation values,
\begin{equation}
\left< \vec\sigma_f \cdot \vec\sigma_N \right>_{S_{1/2},F = 1} = 1 \,,
\qquad
\left< \vec\sigma_f \cdot \vec\sigma_N \right>_{S_{1/2},F = 0} = -3 \,.
\end{equation}
%
Expressed as a relative correction to the leading term,
given in Eq.~\eqref{HSLEAD}, one has
the following corrections due to the X17 particle,
\begin{subequations}
\label{CORRnS12}
\begin{align}
\frac{E_{X,V}(nS_{1/2}) }{ E_F(nS_{1/2}) } 
\approx & \; -\frac{2 \calh'_f \calh'_N}{\gg_N \pi} \, \frac{Z \,m_r}{m_X} \,,
\\[0.1133ex]
\frac{E_{X,A}(nS_{1/2}) }{ E_F(nS_{1/2}) } 
\approx & \; \frac{\calh_f \calh_N}{\gg_N \pi} \, \frac{Z \,m_r}{m_X} \,,
\end{align}
\end{subequations}
depending on the vector ($V$) or pseudoscalar ($A$) hypothesis.
One notices the different sign of the correction,
depending on the symmetry group of the new particle.
We observe that the {\em relative} correction to the 
Fermi splitting is enhanced for muonic bound systems,
by a factor $m_r/m_X \sim m_\mu/m_X$ as compared 
to electronic bound systems, 
because the corresponding factor $m_e/m_X$ is two orders of 
magnitude smaller.

For $nP_{1/2}$ states, whose wave function 
vanishes at the nucleus in the nonrelativistic approximation, 
one finds for the first-order corrections
$E_{X,V}(nP_{1/2}) = 
\left< nP_{1/2} | H_{{\rm HFS},V} | nP_{1/2} \right>$ and
$E_{X,A}(nP_{1/2}) = 
\left< nP_{1/2} | H_{{\rm HFS},A} | nP_{1/2} \right>$,
\begin{subequations}
\begin{align}
\label{EXVnP12}
E_{X,V}(nP_{1/2}) = & \;
\calh'_f \, \calh'_N \, 
\frac{(Z\alpha)^5}{\pi n^3} \, 
\left( 1 - \frac{1}{n^2} \right) 
\nonumber\\[0.1133ex]
& \; \times \frac{m_r^5}{m_f \, m_N \, m_X^2} 
\left< \vec\sigma_f \cdot \vec\sigma_N \right>_{nP_{1/2},F} \,,
\\[0.1133ex]
\label{EXAnP12}
E_{X,A}(nP_{1/2}) = & \;
\calh_f \, \calh_N \, 
\frac{(Z\alpha)^5}{2 \pi n^3} \, 
\left( 1 - \frac{1}{n^2} \right) 
\nonumber\\[0.1133ex]
& \; \times 
\frac{m_r^5}{m_f \, m_N \, m_X^2} 
\left< \vec\sigma_f \cdot \vec\sigma_N \right>_{nP_{1/2},F} \,.
\end{align}
\end{subequations}
In these results, matrix elements of 
tensor structures proportional to 
$\langle \vec\sigma_f \cdot \vec r \; \vec\sigma_N \cdot \vec r \rangle$
in Eqs.~\eqref{HHFS1} and~\eqref{HHFS2}
have been reduced to simpler structures
$\langle \vec\sigma_f \cdot \vec\sigma_N \rangle$
by angular algebra reduction formulas,
which are familar in atomic physics~\cite{VaMoKh1988}.
Under the replacement $\calh'_f \to \calh_f$ 
and $\calh'_N \to \calh_N$,
the correction, for a vector X17 particle,
assumes the same form as for the pseudoscalar hypothesis,
up to an additional overall factor $1/2$.
For $nP_{1/2}$ states, the expectation values are
\begin{equation}
\left< \vec\sigma_f \cdot \vec\sigma_N \right>_{P_{1/2},F = 1} = -\frac13 \,,
\qquad
\left< \vec\sigma_f \cdot \vec\sigma_N \right>_{P_{1/2},F = 0} = 1 \,.
\end{equation}
The leading term in the hyperfine splitting for $nP_{1/2}$ 
states is well known to be equal to 
\begin{align}
E_F(nP_{1/2}) =& \; \left< nP_{1/2}) | H_F | nP_{1/2} \right> 
\nonumber\\[0.1133ex]
=& \; -\gg_N \, \frac{\alpha \, 
(Z\alpha)^3 \, m_r^3}{3 \, n^3 \, m_f \, m_N} \,
\left< \vec\sigma_f \cdot \vec\sigma_N \right>_{nP_{1/2},F} \,.
\end{align}
Expressed in terms of the leading term,
one obtains the following corrections due to the X17 particle
for $nP_{3/2}$ states,
\begin{subequations}
\begin{align}
\label{CORRVnP12}
\frac{E_{X,V}(nP_{1/2}) }{ E_F(nP_{1/2}) } 
\approx & \;
-\frac{3 \calh'_f \calh'_N}{\gg_N \pi} \, \frac{Z \,m_r}{m_X} 
\, \left( 1- \frac{1}{n^2} \right) \,
\left( \frac{Z \alpha m_r}{m_X} \right) \,,
\\[0.1133ex]
\label{CORRAnP12}
\frac{E_{X,A}(nP_{1/2}) }{ E_F(nP_{1/2}) } 
\approx & \;
-\frac{3 \calh_f \calh_N}{2 \gg_N \pi} \, \frac{Z \,m_r}{m_X} 
\, \left( 1- \frac{1}{n^2} \right) \,
\left( \frac{Z \alpha m_r}{m_X} \right) \,.
\end{align}
\end{subequations}
Parametrically, these are suppressed with respect to the 
results for $S$ states, by an additional factor $Z \alpha m_r/m_X$.
For the $nP_{3/2}$ states, one considers the corrections
$E_{X,V}(nP_{3/2}) = \left< nP_{3/2} | H_{{\rm HFS},V} | nP_{3/2} \right>$
and 
$E_{X,A}(nP_{3/2}) = \left< nP_{3/2} | H_{{\rm HFS},A} | nP_{3/2} \right>$,
with the results
\begin{subequations}
\begin{align}
\label{EXVnP32}
E_{X,V}(nP_{3/2}) = & \;
-\frac{(Z\alpha)^5}{12 \pi n^3} \,
\left( 1 - \frac{1}{n^2} \right) 
\frac{m_r^5}{m_N^2 \, m_X^2}
\nonumber\\[0.1133ex]
& \; \times 
\calh'_f \, \calh'_N \,
\left< \vec\sigma_f \cdot \vec\sigma_N \right>_{nP_{3/2},V} \,,
\\[0.1133ex]
\label{EXAnP32}
E_{X,A}(nP_{3/2}) = & \;
\frac{2 (Z\alpha)^6}{45 \pi n^3} \,
\left( 1 - \frac{1}{n^2} \right) \,
\frac{m_r^6}{m_f \, m_N \, m_X^3}
\nonumber\\[0.1133ex]
& \; \times 
\calh_f \, \calh_N \,
\left< \vec\sigma_f \cdot \vec\sigma_N \right>_{nP_{3/2},A}  \,.
\end{align}
\end{subequations}
Here, the expectation values are
\begin{equation}
\left< \vec\sigma_f \cdot \vec\sigma_N \right>_{P_{3/2},F = 2} = 1  \,,
\qquad
\left< \vec\sigma_f \cdot \vec\sigma_N \right>_{P_{3/2},F = 1} = -\frac53 \,.
\end{equation}
The leading term in the hyperfine splitting for $nP_{3/2}$
states is well known to be equal to
\begin{align}
E_F(nP_{3/2}) =& \; \left< nP_{3/2}) | H_F | nP_{3/2} \right> 
\nonumber\\[0.1133ex]
=& \; \gg_N \, \frac{\alpha \, 
(Z\alpha)^3 \, m_r^3}{15 \, n^3 \, m_f \, m_N} \,
\left< \vec\sigma_f \cdot \vec\sigma_N \right>_{nP_{3/2}} \,.
\end{align}
Expressed in terms of the leading term,
one obtains the following corrections due to the X17 particle
for $nP_{1/2}$ states,
\begin{subequations}
\begin{align}
\label{CORRVnP32}
\frac{E_{X,V}(nP_{3/2}) }{ E_F(nP_{3/2}) } 
\approx & \;
-\frac{5 \calh'_f \calh'_N}{4 \gg_N \pi} \, \frac{Z \,m_r}{m_X} \,
\frac{m_f}{m_N} \,
\nonumber\\[0.1133ex]
& \; \times \left( 1- \frac{1}{n^2} \right) \,
\left( \frac{Z \alpha m_r}{m_X} \right) \,,
\\[0.1133ex]
\label{CORRAnP32}
\frac{E_{X,A}(nP_{3/2}) }{ E_F(nP_{3/2}) } 
\approx & \;
\frac{2 \calh_f \calh_N}{3 \gg_N \pi} \, \frac{Z \,m_r}{m_X} 
\, \left( 1- \frac{1}{n^2} \right) \,
\left( \frac{Z \alpha m_r}{m_X} \right)^2 \,.
\end{align}
\end{subequations}
Parametrically, in comparison to $nP_{1/2}$ states,
the correction for $nP_{3/2}$ states 
is suppressed for a vector X17 by an additional
factor $m_f/m_N$, while for a pseudoscalar X17,
the suppression factor is $Z \alpha m_r/m_X$.
For electrons bound to protons and other nuclei,
both suppression factors are approximately
of the same order-of-magnitude, while
for muonic hydrogen and deuterium, the vector contribution 
dominates over the pseudoscalar one.

\begin{figure}[t!]
\begin{center}
\begin{minipage}{0.99\linewidth}
\begin{center}
\includegraphics[width=0.91\linewidth]{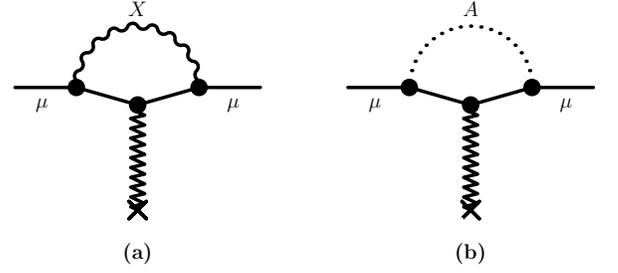}
\caption{\label{fig2} The X17 particle induces
vertex corrections to the anomalous magnetic 
moment of the muon. For the X17 vector hypothesis,
one obtains diagram (a), while the pseudoscalar 
hypothesis leads to diagram (b).
The interaction with the external magnetic field
is denoted by a zigzag line.}
\end{center}
\end{minipage}
\end{center}
\end{figure}

\begin{figure}[t!]
\begin{center}
\begin{minipage}{0.99\linewidth}
\begin{center}
\includegraphics[width=0.91\linewidth]{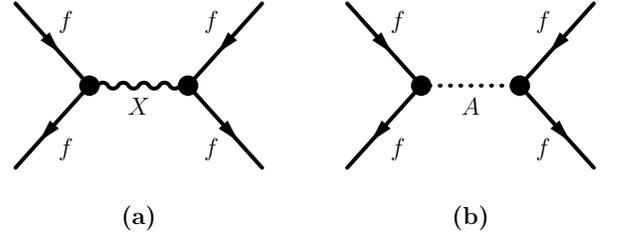}
\caption{\label{fig3} 
For bound systems 
consisting of a lepton and anti-lepton
there is an additional correction to the 
energy levels induced by virtual annihilation
(the arrow of time is from left to right).
We here consider $f=e$ (positronium)
and $f=\mu$ (true muonium).
The resulting effective potential is proportional
to a Dirac-$\delta$ and affects $S$ states.
Diagram (a) is relevant for orthopositronium and
ortho true muonion ($S=1$, annihilation into a vector X17 boson)
while diagram (b) is relevant for para states
(with total spin
$S=0$, which allows for an 
annhilation into a hypothetical pseudoscalar X17 boson).
Both virtual processes contribute to the 
hyperfine splitting.}
\end{center}
\end{minipage}
\end{center}
\end{figure}

%
%
\subsection{Virtual Annihilation}

For bound systems consisting of a particle
and antiparticle, virtual annihilation processes
also need to be considered (see Fig.~\ref{fig3}).
The resulting effective potentials are
local (proportional to a Dirac-$\delta$)
and affect $S$ states.
We here consider positronium 
and true muonium ($f = e,\mu$), for which one 
has $Z=1$, $N = \overline f$ (antifermion), $m_r = m_f/2$,
$m_N = m_f$. 
In this case, the Fermi energy, as defined in Eq.~\eqref{EFERMI}, 
assumes the form (it is no longer equal to the full leading-order result
for the hyperfine splitting, as we will see)"
\begin{equation}
E_{F,f{\overline f}}(nS) = \frac{\alpha^4 \, m_f}{12 \, n^3} \,
\langle \vec\sigma_f \cdot \vec\sigma_{\overline f} \rangle \,.
\end{equation}
When one replaces the vector X17 boson in 
Fig.~\ref{fig3}(a) by a photon, one obtains
the annihilation potential [see Eq.~(83.24) of 
Ref.~\cite{BeLiPi1982vol4}],
\begin{equation}
\label{HANNgamma}
H_{{\rm ANN},\gamma} =
\frac{\pi \alpha}{2 m_f^2} 
\left( \vec\sigma_f \cdot \vec\sigma_{\overline f} + 3 \right) \,
\delta^{(3)}(\vec r) \,.
\end{equation}
Based on the identity
\begin{equation}
\langle \vec\sigma_f \cdot \vec\sigma_{\overline f} \rangle = 
2 S (S+1) -3 \,,
\end{equation}
where $S=1$ for an ortho state, and $S = 0$ for a para state,
one can see that the annihilation process
into a virtual vector particle is relevant only  
for ortho states, consistent with the conservation
of total angular momentum in the virtual transition
to the photon, which has an intrinsic spin of unity.
The virtual annihilation contribution 
to the hyperfine splitting is
\begin{equation}
E_{{\rm ANN},\gamma}(nS) = \frac{\alpha^4 \, m_f}{16 \, n^3} \,
\langle \vec\sigma_f \cdot \vec\sigma_{\overline f} \rangle \,.
\end{equation}
The difference between the expectation 
values for ortho and para states is the 
well-known result
[$\langle\langle
\vec\sigma_f \cdot \vec\sigma_{\overline f} \rangle\rangle = 
1 - (-3) = 4$]
\begin{equation}
\Delta E_{\rm HFS}(nS) = 
\langle\langle
E_{F}(nS) +
E_{{\rm ANN},\gamma}(nS)
\rangle\rangle
= \frac{7}{12} \frac{\alpha^4 \, m_f}{n^3} \,.
\end{equation}
The exchange of a virtual photon contributes a fraction
of $4/7$ to this result, while the virtual
annihilation yields the remaining fraction of $3/7$.

The generalization of Eq.~\eqref{HANNgamma}
to a vector X17 exchange is immediate,
\begin{equation}
\label{HANNV}
H_{{\rm ANN}, V} =
\frac{(\calh'_f)^2}{8 (m_f^2 - \tfrac14 m_X^2)} 
\left( \vec\sigma_f \cdot \vec\sigma_{\overline f} + 3 \right) 
\delta^{(3)}(\vec r) \,,
\end{equation}
with the expectation value (we select the 
term relevant to the hyperfine splitting)
\begin{equation}
\label{EANNV}
E_{{\rm ANN}, V}(nS) =
\frac{(\calh'_f)^2 \, (\alpha m_f)^3}{64 (m_f^2 - \tfrac14 m_X^2) n^3} 
\langle \vec\sigma_f \cdot \vec\sigma_{\overline f} \rangle \,.
\end{equation}
In the calculation, one precisely follows
Eqs.~(83.18)--(83.24) of Ref.~\cite{BeLiPi1982vol4}
and adjusts for the mass of the X17 in the propagator
denominator.
The relative correction to the hyperfine splitting 
due to annihilation channel, for a virtual vector X17 particle
(for $S$ states), is 
\begin{equation}
\label{rEANNV}
\frac{E_{{\rm ANN},V}(nS)}{E_{{\rm ANN},\gamma}(nS)} =
\frac{(\calh'_f)^2}{4 \pi \alpha} \frac{m_f^2}{m_f^2 - \tfrac14 m_X^2} \,.
\end{equation}

For the virtual annihilation into a 
pseudoscalar particle, one can also 
follow Eqs.~(83.18)--(83.24) of Ref.~\cite{BeLiPi1982vol4},
but one has to adjust for the 
different interaction Lagrangian, 
which now involves the fifth current,
and one also needs to adjust for the mass of the X17 in the pseudoscalar
propagator denominator. 
The result in Eq.~(83.22) of Ref.~\cite{BeLiPi1982vol4}
for the Fierz transformation of the 
currents has to be adapted to the 
pseudoscalar current, i.e., to the last 
entry in Eq.~(28.17) of Ref.~\cite{BeLiPi1982vol4}.
The result is
\begin{equation}
\label{HANNA}
H_{{\rm ANN}, A} =
\frac{3 \calh_f^2}{8 (m_f^2 - \tfrac14 m_X^2)} 
\left( \vec\sigma_f \cdot \vec\sigma_{\overline f} - 1 \right) 
\delta^{(3)}(\vec r) \,.
\end{equation}
The expectation value of this effective Hamiltonian
is nonvanishing only for 
para states ($S=0$), as had to be expected 
in view of the pseudoscalar nature of the 
virtual particle (the intrinsic parity of 
para states of positronium and true muonium 
is negative, allowing for the virtual 
transition to the pseudoscalar X17).
In contrast to Eq.~\eqref{HHFS2}, we observe a
small shift of the hyperfine centroid for the fermion-antifermion
system, due to the term that is added to the scalar product of the
spin operators.
The general expression for the expection value 
in and $nS$ state is (we select the
term relevant to the hyperfine splitting)
\begin{equation}
\label{EANNA}
E_{{\rm ANN}, A}(nS) =
\frac{3 \calh_f^2 (\alpha m_f)^3}{64 (m_f^2 - \tfrac14 m_X^2) n^3} 
\langle \vec\sigma_f \cdot \vec\sigma_{\overline f} \rangle \,.
\end{equation}
The relative correction to the hyperfine splitting due to the 
pseudoscalar annihilation channel, for $S$ states, is  
\begin{equation}
\label{rEANNA}
\frac{E_{{\rm ANN},A}(nS)}{E_{{\rm ANN},\gamma}(nS)} =
\frac{3 (\calh_f)^2}{4 \pi \alpha} \frac{m_f^2}{m_f^2 - m_X^2} \,.
\end{equation}
For the fermion-antifermion bound system, the corrections
given in Eq.~\eqref{CORRnS12} specialize as follows,
\begin{subequations}
\label{rEEXCH}
\begin{align}
\frac{E_{X,V}(nS) }{ E_F(nS) }
\approx & \; -\frac{(\calh'_f)^2}{2 \pi} \, \frac{m_f}{m_X} \,,
\\[0.1133ex]
\frac{E_{X,A}(nS) }{ E_F(nS) }
\approx & \; \frac{(\calh_f)^2}{4 \pi} \, \frac{m_f}{m_X} \,,
\end{align}
\end{subequations}
The relative correction to the total hyperfine splitting, 
due to the X17 exchange and annihilation channels, is 
\begin{subequations}
\label{defchi}
\begin{align}
\chi_V(nS) = & \;
\frac47 \frac{E_{X,V}(nS) }{ E_F(nS) } +
\frac37 \frac{E_{{\rm ANN},V}(nS)}{E_{{\rm ANN},\gamma}(nS)}  \,,
\\[0.1133ex]
\chi_A(nS) = & \;
\frac47 \frac{E_{X,A}(nS) }{ E_F(nS) } +
\frac37 \frac{E_{{\rm ANN},A}(nS)}{E_{{\rm ANN},\gamma}(nS)}  \,,
\end{align}
\end{subequations}
with the individual contributions given in
Eqs.~\eqref{CORRnS12},~\eqref{rEANNV},~\eqref{rEANNA} and~\eqref{rEEXCH}.

%
%
\section{X17 Particle and Muon Anomalous Magnetic Moment}
\label{sec5}

One aim of our investigations
is to explore the possibility of a detection
of the X17 particle in the hyperfine splitting
of muonic bound systems.
To this end, it is instructive to derive upper
bounds on the coupling parameters $\calh'_\mu$ 
and $\calh_\mu$, for the muon.
The contribution of a massive pseudoscalar 
loop to the muon anomaly [see Fig.~\ref{fig2}(b)] has been studied
for a long 
time~\cite{BaGaLa1972,PrQu1972,Le1978npb,HaKaSt1979,QuSh2014},
and recent updates of theoretical 
contributions~\cite{DaHoMaZh2017,KeNoTe2018,DaHoMaZh2020}
has confirmed the existence of a (roughly) $3.5$ discrepancy
of theory and experiment. The contribution
of a massive vector exchange [see Fig.~\ref{fig2}(a)] has recently
been revisited in Ref.~\cite{QuSh2014}. Specifically, 
the experimental results for the muon anomaly
$a_\mu$ [see Eqs.~(1.1) and (3.36) of Ref.~\cite{KeNoTe2018}]
are as follows,
\begin{align}
\label{disc}
a^{\rm (exp)}_\mu =& \; 0.00116592091(54)(33) \,,
\\[0.1133ex]
a^{\rm (thr)}_\mu =& \; 0.001165918204(356)\,.
\end{align}
The $3.7\sigma$ discrepancy $a^{\rm (exp)}_\mu - 
a^{\rm (thr)}_\mu \approx 2.7 \times 10^{-9}$ 
needs to be explained.

According to  Eq.~(41) of Ref.~\cite{QuSh2014}, we have
the following correction to the muon anomaly
due to the vector $X$ vertex correction 
in Fig.~\ref{fig2}(a),
\begin{align}
\label{DeltaG_V}
\Delta a_\mu = & \;
\frac{(\calh'_\mu)^2}{8 \pi^2} 
\frac{m_\mu^2}{m_X^2} \,
\int\limits_0^1 \frac{\dd x \, x^2 \, (2-x)}{(1-x) \, 
\left[1 - \frac{m^2_\mu}{m^2_X}\right] +
\frac{m^2_\mu}{m^2_X} \, x} 
\nonumber\\[0.1133ex]
=& \; 8.64 \times 10^{-3} \, (\calh'_\mu)^2 \,,
\end{align}
where we have used the numerical value
$m_X = 16.7 \, {\rm MeV}$.  The following numerical value
\begin{equation}
\label{hpmuV}
\calh'_\mu = 
(\calh'_\mu)_{\rm opt} = 
5.6 \times 10^{-4}
\end{equation}
is ``optimum'' in the sense that it
precisely remedies the discrepancy described by Eq.~\eqref{disc}
and will be taken as the input datum for all 
subsequent evaluations of corrections to the 
hyperfine splitting in muonic bound systems.
Note that, even if the vector X17 particle 
does not provide for an explanation of the 
muon anomaly discrepancy, 
the order-of-magnitude of the 
coupling parameter $\calh'_\mu$ could not
be larger than the value indicated in 
Eq.~\eqref{hpmuV}, because otherwise, the 
theoretical value of $a_\mu$ would increase 
too much beyond the experimental result.

According to  Eq.~(20) of Ref.~\cite{QuSh2014},
the vertex correction due to a virtual pseudoscalar X17 leads to the 
following correction,
\begin{align}
\label{DeltaG_A}
\Delta a_\mu = & \;
-\frac{(\calh_\mu)^2}{4 \pi^2} 
\frac{m_\mu^2}{m_X^2} \,
\int\limits_0^1 \frac{\dd x \, x^3}{(1-x) \, 
\left[1 - \frac{m^2_\mu}{m_X^2} \right] +
\frac{m_\mu}{m_X}^2 \, x} 
\nonumber\\[0.1133ex]
=& \; -1.19 \times 10^{-3} \, (\calh_\mu)^2 \,.
\end{align}
Here, because the correction is negative and 
decreases the value of $a_\mu$, the experimental-theoretical
discrepancy given in Eq.~\eqref{disc} can only be enhanced
by the pseudoscalar X17 particle.
If we demand that the discrepancy not be increased
beyond $6 \sigma$, then we obtain the 
condition that $|\calh_\mu|$ could not exceed
a numerical value of $3.8 \times 10^{-4}$.
In the following, we take the maximum permissible value of
\begin{equation}
\label{hmuA}
\calh_\mu = 
(\calh_\mu)_{\rm max} = 
3.8 \times 10^{-4} \,,
\end{equation}
in order to estimate the magnitude of 
corrections to the hyperfine splitting in muonic bound systems,
induced by a hypothetical pseudoscalar X17 particle.

%
%
\section{Numerical Estimates and Experimental Verification}
\label{sec6}

%
%
\subsection{Overview}
\label{sec61}

The relative corrections to the hyperfine splitting
due to the X17 particle, expressed in terms
of the leading Fermi interaction, for $S$ and $P$ states,
are given in Eqs.~\eqref{CORRnS12},~\eqref{CORRVnP12},~\eqref{CORRAnP12},%
~\eqref{CORRVnP32}, and~\eqref{CORRAnP32}.
All of the formulas involve at least one factor 
of $m_r/m_X$, and so, experiments appear to be 
more attractive for muonic rather than electronic 
bound systems.
Furthermore, parametrically, the corrections are largest
for $S$ states, which is understandable because 
the range of the X17 potential is limited to 
its Compton wavelength of about $11.8\,{\rm fm}$,
and so, its effects should be more pronounced for
states whose probability density does not vanish 
at the origin, i.e., for $S$ states.
This intuitive understanding is confirmed by 
our calculations. Note that the
formulas for the corrections to the 
hyperfine splitting, given in Eqs.~\eqref{EXVnS12},~\eqref{EXAnS12},%
~\eqref{EXVnP12},~\eqref{EXAnP12},%
~\eqref{EXVnP32}, and~\eqref{EXAnP32},
are generally applicable to bound systems
with a heavy nucleus, upon a suitable 
reinterpretation of the nuclear spin matrix
$\vec\sigma_N$ in terms of a nuclear 
spin operator.

%
%
\subsection{Muonic Deuterium}
\label{sec62}

In view of a recent theoretical work~\cite{KaPaYe2018}
which describes a $5 \sigma$ discrepancy of 
theory and experiment for muonic deuterium,
it appears indicated to analyze this system first.
Indeed, the theory of nuclear-structure effects
in muonic deuterium has been updated a 
number of times in recent years~\cite{Pa2011,PaWi2015,KaPaYe2018}.
Expressed in terms of the Fermi term, the discrepancy 
$\delta E_{\rm HFS}(2S)$ observed in 
Ref.~\cite{KaPaYe2018} can be written as
\begin{equation}
\label{discrep}
\frac{\Delta E_{\rm HFS}(2S_{1/2})}{E_F(2S_{1/2})} 
= 0.0094(18) \,.
\end{equation}
In order to evaluate an estimate for the correction 
due to the X17 vector particle, we observe that the 
interaction is protophobic~\cite{FeEtAl2016,FeEtAl2017}. Hence, we can 
assume that the coupling parameter of the deuteron 
is approximately equal to that of the neutron.
We will assume the opposite sign for the 
coupling parameter of the deuteron (neutron),
as compared to the sign of the coupling parameter in Eq.~\eqref{hpmuV}.
This choice is inspired
by the opposite charge of the muon and nucleus with respect to the
$U(1)$ gauge group of quantum electrodynamics.
In view of Eq.~(10) of Ref.~\cite{FeEtAl2016}
and Eq.~\eqref{valuen} here, we thus have the estimate
\begin{equation}
\calh'_d \approx
\calh'_n = -\frac{1}{100} \, \sqrt{4 \pi \alpha} =
- 3.02 \times 10^{-3} \,.
\end{equation}
Because of the numerical dominance of the proton 
coupling to the pseudoscalar particle
over that of the neutron [see Eq.~\eqref{hpn}], we 
estimate the pseudoscalar coupling of the deuteron
to be of the order of
\begin{equation}
\calh_d \approx
\calh_p = - 2.4 \times 10^{-3} \,.
\end{equation}
In view of Eq.~\eqref{CORRnS12},
we obtain the estimates
\begin{subequations}
\begin{align}
\frac{ E^{(\mu d)}_{X,V}(nS_{1/2}) }{ E_F(nS_{1/2}) }
\vapprox & \;
3.8 \times 10^{-6} \,,
\\[0.1133ex]
\frac{ E^{(\mu d)}_{X,A}(nS_{1/2}) }{ E_F(nS_{1/2}) }
\vapprox & \;
-1.0 \times 10^{-6} \,,
\end{align}
\end{subequations}
where the symbol $\vapprox$ is used to denote an
estimate for the quantity specified on the left, 
including its sign, based on the estimates of the 
coupling parameters of the hypothetical vector 
and pseudoscalar X17 particle, as described in the current work.
As already explained, the modulus of 
our estimates for the coupling parameters 
is close to the upper end of the allowed range;
the same thus applies to the absolute magnitude of our 
estimates for the X17-mediated corrections to hyperfine energies.
Note that the vector X17 contribution slightly decreases the 
discrepancy noted in Ref.~\cite{KaPaYe2018}, while
the hypothetical pseudoscalar effect slightly increases the 
discrepancy, yet, on a numerically almost negligible level.

Similar considerations, based on Eqs.~\eqref{CORRVnP12} and~\eqref{CORRVnP12},
lead to the following results for $P$ states,
\begin{subequations}
\label{muD_results}
\begin{align}
\frac{E^{(\mu d)}_{X,V}(nP_{1/2}) }{ E_F(nP_{1/2}) }
\vapprox & \; 2.5 \times 10^{-7} \, \left( 1 - \frac{1}{n^2} \right) \,,
\\[0.1133ex]
\frac{E^{(\mu d)}_{X,A}(nP_{1/2}) }{ E_F(nP_{1/2}) }
\vapprox & \; 6.6 \times 10^{-8} \, \left( 1 - \frac{1}{n^2} \right) \,,
\end{align}
\end{subequations}
which might be measurable in future experiments.
Specifically, there is a nuclear-structure 
correction to the $P_{1/2}$ state hyperfine
splitting due to the lower component of the 
Dirac $nP_{1/2}$ wave function, which has 
$S$-state symmetry. However, the lower component
of the wave function is suppressed by a factor $\alpha$,
which implies that the $P_{1/2}$ state 
nuclear-structure correction is suppressed 
in relation to $E_F(nP_{1/2})$ by a factor $\alpha^2$.
An order-of-magnitude of the achievable theoretical 
uncertainty for the $P_{1/2}$-state result 
can be given by an appropriate scaling of the 
current theoretical uncertainty 
for $S$ states, given in Eq.~\eqref{discrep}.
The result is that the achievable theoretical uncertainty
should be better than $10^{-7}$,
which would make the effect given in Eq.~\eqref{muD_results}
measurable. Also, according to Ref.~\cite{PoPriv2020},
the experimental accuracy should be 
improved into the range of $10^{-6} \dots 10^{-7}$ in the
next round of planned experiments.
Results for the Sternheim~\cite{St1963} weighted 
differences $[n^3 E_{X,V}(nS_{1/2}) - E_{X,V}(1S_{1/2})]/E_F(1S_{1/2})$
and correspondingly 
$[n^3 E_{X,A}(nS_{1/2}) - E_{X,A}(1S_{1/2})]/E_F(1S_{1/2})$
are of the same order-of-magnitude
as for the individual $P_{1/2}$ states.

%
%
\subsection{Muonic Hydrogen}
\label{sec63}

The considerations are analogous to those for muonic deuterium.
However, the coupling parameter for the nucleus,
for the protophobic vector model, is constrained by Eq.~\eqref{constrainp},
\begin{equation}
\calh'_p \approx
 -8 \times 10^{-4} \, \sqrt{4 \pi \alpha} =
- 2.42 \times 10^{-5} \,,
\end{equation}
which is much smaller than for the deuteron nucleus.
The coupling parameter of the proton, for the pseudoscalar 
model, can be estimated according to Eq.~\eqref{hp}.
One obtains
\begin{subequations}
\begin{align}
\frac{E^{(\mu p)}_{X,V}(nS_{1/2}) }{ E_F(nS_{1/2}) }
\vapprox & \; 8.8 \times 10^{-9} \,,
\\[0.1133ex]
\frac{E^{(\mu p)}_{X,A}(nS_{1/2}) }{ E_F(nS_{1/2}) }
\vapprox & \; -2.9 \times 10^{-7} \,,
\end{align}
\end{subequations}
for the $S$ state effects, and 
\begin{subequations}
\begin{align}
\frac{E^{(\mu p)}_{X,V}(nP_{1/2}) }{ E_F(nP_{1/2}) }
\vapprox & \; 5.8 \times 10^{-10} \, \left( 1 - \frac{1}{n^2} \right) \,,
\\[0.1133ex]
\frac{E^{(\mu p)}_{X,A}(nP_{1/2}) }{ E_F(nP_{1/2}) }
\vapprox & \; 1.8 \times 10^{-8} \, \left( 1 - \frac{1}{n^2} \right) \,,
\end{align}
\end{subequations}
for $P_{1/2}$ states.
Results of the same order-of-magnitude 
as for individual $P_{1/2}$ states are obtained
for the Sternheim difference
of $S$ states.
The effects, in muonic hydrogen, for the vector model,
are seen to be numerically suppressed.
The contributions of the X17 particle need to be
compared to the proton structure effects,
which have recently been analyzed in 
Refs.~\cite{Ma2005mup,PoGiMiPa2011,To2017,To2019epja1,To2019epja2}.
According to Ref.~\cite{To2019epja1}, the 
numerical accuracy of the theoretical prediction
for the $2S$ hyperfine splitting in muonic hydrogen
is currently about $72\,$ppm
[$E_{\rm HFS}(2S) = 22.8108(16) \, {\rm meV}$].

%
%
\subsection{Muonium}

Muonium is the bound system consisting of a 
positively charged antimuon ($\mu^+$), 
and an electron ($e^-$).
Its ground-state hyperfine splitting has been studied in
Ref.~\cite{LiEtAl1999} with a 
result of $\Delta \nu^{\rm (exp)}_{\rm HFS} = 4\,463\,302\,765(53) \, {\rm Hz}$,
i.e., with an accuracy of $1.2 \times 10^{-8}$.
The theoretical uncertainty is about one order-of-magnitude worse
and amounts to $1.2 \times 10^{-7}$ (see Ref.~\cite{Ei2019}),
with the current status being summarized in the theoretical
prediction
$\Delta \nu^{\rm (thr)}_{\rm HFS} = 4\,463\,302\,872(515)\,{\rm Hz}$.

Coupling parameters for the muon have been estimated
in Eqs.~\eqref{hpmuV} and~\eqref{hmuA} for the
vector and pseudoscalar models, respectively,
and we take the antimuon coupling estimate
as the negative value of the coupling parameters for the 
muon.
For the coupling parameters, we use the
maximum allowed value for the electron in the
vector model [see Eq.~\eqref{epsEbound}],
\begin{equation}
\calh'_e \vapprox 1.4 \times 10^{-3} \, \sqrt{4 \pi \alpha} =
4.2 \times 10^{-4} \,,
\end{equation}
and for the pseudoscalar model [see Eq.~\eqref{xi_e_upper}],
\begin{equation}
\calh_e \vapprox 500 \, \frac{m_e}{v} = 1.0 \times 10^{-3} \,.
\end{equation}
One obtains the estimates
\begin{subequations}
\begin{align}
\frac{E^{(\mu \mu)}_{X,V}(nS_{1/2}) }{ E_F(nS_{1/2}) }
\vapprox & \; 2.3 \times 10^{-9} \,,
\\[0.1133ex]
\frac{E^{(\mu \mu)}_{X,A}(nS_{1/2}) }{ E_F(nS_{1/2}) }
\vapprox & \; -1.8 \times 10^{-9} \,.
\end{align}
\end{subequations}
Because the reduced mass of muonium is 
close to the electron mass,
the effect of the X17 boson is parametrically suppressed.
It will take considerable effort to increase experimental 
precision beyond the level attained in 
Ref.~\cite{LiEtAl1999}.
On the other hand, hadronic vacuum polarization effects
are suppressed in muonium, and their
uncertainty~\cite{KaSh2001} is less than
the X17-induced effect 
in the hyperfine splitting of muonium.
It is thus not completely hopeless to detect 
X17-induced effects in muonium in the future.

%
%
\subsection{True Muonium (\texorpdfstring{$\maybebm{\mu^+\mu^-}$}{mu+ mu-}) System}
\label{sec64}

Taking into account the exchange and annihilation channels,
and using the same coupling parameter estimates
as for muonium, one obtains the estimates
[see Eq.~\eqref{defchi}]
\begin{equation}
\chi_V(nS) \vapprox 1.3 \times 10^{-6} \,,
\qquad
\chi_A(nS) \vapprox 2.1 \times 10^{-6} \,.
\end{equation}
The annihilation channel contribution numerically 
dominates over the exchange channel.
For $S$ states, the contribution of hadronic vacuum 
polarization in the annihilation channel
has been improved to the 
level of $2\,$ppm, as a result of gradual 
progress achieved over the last decades
[see Eq.~(41) of Ref.~\cite{OwRe1972,JeSoIvKa1997}
as well as Refs.~\cite{OwRe1972,KaJeIvSo1998,BrLe2009},
and the recent work~\cite{La2017tm}].
A very modest progress in the theoretical determination
of the hadronic vacuum-polarization contribution
would make the effect of the X17 visible.



%
%
\subsection{Positronium}
\label{sec65}

Quite considerable efforts have recently 
been invested in the calculation of the 
$m \alpha^7$ corrections to the positronium
hyperfine splitting,
and related effects
\cite{PePi2017,HeWiRuCr2018,CzMeYe1999prl,CzMeYe1999pra,BaEtAl2014,AdFe2014,%
EiSh2014,AdEtAl2014,EiSh2015,AdPaSaWa2015,%
AdKiPaFe2015,AdTaWa2016,EiSh2016,EiSh2017}.
In positronium,
effects of the X17 particle are suppressed in view of the smaller
reduced mass of the bound system.
Under these assumptions, the estimates for $S$ states are as follows,
[see Eq.~\eqref{defchi}]
\begin{equation}
\chi_V(nS) \vapprox -3.6 \times 10^{-9} \,,
\qquad
\chi_A(nS) \vapprox -5.1 \times 10^{-9} \,.
\end{equation}
The effects are thus numerically smaller than the 
$m \alpha^7$ effects currently under 
study~\cite{HeWiRuCr2018,CzMeYe1999prl,CzMeYe1999pra,BaEtAl2014,AdFe2014,%
EiSh2014,AdEtAl2014,EiSh2015,AdPaSaWa2015,%
AdKiPaFe2015,AdTaWa2016,EiSh2016,EiSh2017}.

%
%
\section{Conclusions}
\label{sec7}

The conceivable existence of the 
X17 particle~\cite{KrEtAl2016,KrEtAl2017woc2,KrEtAl2019}
provides atomic physicists with 
a long-awaited opportunity to detect
a very serious candidate for a low-energy
(fifth force) addition to the Standard Model.
The energy range of about $17$\,MeV provides 
for a certain challenge from the viewpoint 
of atomic physics;
the range of the X17-induced interaction 
potentials is smaller than the 
effective Bohr radii in 
electronic and muonic bound systems.
Rather than looking at the Lamb shift~\cite{JeNa2018},
we here advocate a closer look at the 
effects induced by the X17 on the hyperfine 
splitting, for both $S$ and $P$ states,
in simple electronic and muonic bound systems.
This notion is based on two observations:
{\em (i)} Hyperfine effects for $S$ states
are induced by a contact interaction
(the Fermi contact term), and thus, 
naturally confined to a distance range 
very close to the atomic nucleus.
As far as hyperfine effects are concerned,
the virtual exchange of an X17 is 
thus not masked by its small range.
{\em (ii)} In the pseudoscalar model~\cite{ElMo2016},
the one-quantum exchange of an X17 exclusively 
leads to hyperfine effects, but leaves the 
Lamb shift invariant.
or a fermion-antifermion
system, this statement should be taken {\em with a small grain
of salt}, namely, it holds up to the numerically tiny shift of
the hyperfine centroid, induced by the virtual annihilation
channel [see Eq.~\eqref{HANNA}].
This finding could be 
of interest irrespective of whether the 
results of the experimental observations
reported in Refs.~\cite{KrEtAl2016,KrEtAl2017woc2,KrEtAl2019} 
can be independently confirmed by other groups.

We have derived limits on the 
coupling parameters of the X17 particle
in the muonic sector in Sec.~\ref{sec5} 
and compiled estimates for the X17-induced effects
in Sec.~\ref{sec6}, for a number of 
simple atomic systems.
The results can be summarized as follows.
\begin{itemize}
\item
We show in Sec.~\ref{sec5} that 
the pseudoscalar model~\cite{ElMo2016} 
enhances the experimental-theoretical discrepancy 
in the muon anomaly, while the vector model~\cite{FeEtAl2016,FeEtAl2017})
could eliminate it.
Stringent bounds on the magnitude of the 
muon coupling parameters to the X17 particle
can be derived based on the muon anomaly.
Note that the order-of-magnitude of the 
maximum permissible coupling to the pseudoscalar,
given in Eq.~\eqref{hmuA}, also leads to a tension
with the parameterization $h_f = \xi_f (m_f/v)$ 
given in Ref.~\cite{ElMo2016}
(applied to the case $f=\mu$, i.e., to the muon).
Namely, the parameterization could be read
as suggesting a likely increase of the 
pseudoscalar coupling parameter with the mass 
of the particle. While $\xi_e$ is bound from below
by the condition $\xi_e > 4$ [see Eq.~\eqref{xi_e_lower}]
the corresponding parameter in the muon sector 
must fulfill $\xi_\mu < 0.9$ [see Eq.~\eqref{hmuA}].
\item The relative correction to the 
hyperfine splitting for both $S$ and $P$ states
is enhanced in muonic 
as compared to electronic bound systems 
by two orders of magnitude, 
in view of the scaling of the relative
corrections with $m_r/m_X$, where $m_r$ is the 
reduced mass of the two-body bound system.
\item In muonic deuterium,
the correction, for $S$ states,
is of order $3.8 \times 10^{-6}$ (vector X17)
and $-1.0 \times 10^{-6}$ (pseudoscalar X17)
in units of the Fermi energy, while the 
experimental accuracy for the $S$ state 
hyperfine splitting is of order 
$10^{-3}$, and there is a $5 \,\sigma$
discrepancy of theory and experiment,
in view of a recent calculation of the 
nuclear polarizability effects~\cite{KaPaYe2018}.
One concludes that the experimental accuracy
would have to be improved by three
orders of magnitude before the effects of the X17 
become visible, and the understanding of the 
nuclear effects would likewise have to be improved 
by a similar factor.
%
\item In muonic deuterium,
for the hyperfine splitting of $P_{1/2}$ states, the X17-mediated 
correction to the hyperfine splitting is of
order $2.5 \times 10^{-7}$ for the 
vector model, and of order
$6.6 \times 10^{-8}$ for the pseudoscalar model.
These effects are not suppressed by challenging 
nuclear structure effects and could be
measurable in the next round of experiments~\cite{PoPriv2020}.
The same applies to the Sternheim weighted 
difference~\cite{St1963} of the hyperfine splitting of $S$
states, where the effect induced by the X17 
particle is of the same order-of-magnitude as 
for the $P_{1/2}$ splitting.

\item In muonic hydrogen, because
of the protophobic character of the 
vector model, effects of the vector X17 
are suppressed (order $10^{-9}$ for the 
$S$ state splitting and order $10^{-10}$ 
for the $P$ state splitting and the Sternheim 
weighted difference).
For the pseudoscalar model, 
the $S$ state splitting is affected at 
relative order $10^{-7}$, and the 
$P$ state splitting as well as the Sternheim difference
are affected at order $10^{-8}$.
These effects could be measurable in the 
future.

\item For muonium, the situation is not hopeless:
While the effects induced by the X17 shift
the hyperfine splitting only on the level 
of $10^{-9}$, which is two orders of magnitude 
lower than current theoretical predictions~\cite{Ei2019},
we can say that, at least, the uncertainty in the 
theoretical treatment of the hadronic vacuum 
polarization~\cite{KaSh2001} would not impede an
experimental detection of the X17. 
Still, it would seem that more attractive possibilities
exist in muonic systems.

\item For true muonium, one expects 
effects (for the hyperfine splitting of $S$ states)
on the order of $10^{-6}$ for the 
vector X17 model as well as the pseudoscalar model.
These have to be compared to the uncertainty 
from the hadronic vacuum polarization,
which has recently been
improved to the level of 2\,ppm~\cite{La2017tm}.
This implies that a modest progress in the
determination of the $R$ ratio, namely,
the ratio of the cross section of an
electron-positron pair going into hadrons, versus an
electron-positron pair going into muons 
[see Eq.~(9) of Ref.~\cite{La2017tm}],
could render the effect visible in true muonium.
\item For positronium, 
the effects of the X17 are suppressed by the 
small reduced mass of the system.
They are bound not to exceed the level of 
$10^{-9}$ for the vector model and pseudoscalar models.
Thus, even taking into account
all the $m \alpha^7$ corrections currently under
study~\cite{HeWiRuCr2018,CzMeYe1999prl,CzMeYe1999pra,BaEtAl2014,AdFe2014,%
EiSh2014,AdEtAl2014,EiSh2015,AdPaSaWa2015,%
AdKiPaFe2015,AdTaWa2016,EiSh2016,EiSh2017},
the detection of an X17-induced signal in the
hyperfine splitting appears to be extremely 
challenging in positronium.
\end{itemize}
One concludes that the most promising approach
toward a conceivable detection of the X17 in 
high-precision atomic physics experiments
would probably concern the hyperfine splitting 
of $P_{1/2}$ states
in muonic deuterium, and the related Sternheim
difference, where the effects are enhanced
because of the large mass ratio of the reduced
mass of the atomic system to the mass of the X17,
and nuclear structure effects are suppressed
because the $P$ state wave function (as well as the 
weighted difference of the $S$ states) vanishes 
at the origin. Furthermore, very attractive prospects in true 
muonium~\cite{OwRe1972,JeSoIvKa1997,KaJeIvSo1998,BrLe2009,La2017tm}
should not be overlooked.

%
%
\section*{Acknowledgments}

The author acknowledges support from the 
National Science Foundation (Grant PHY--1710856).
Also, the author acknowledges utmost insightful 
conversations with K.~Pachucki.
Helpful conversations with A.~Krasznahorkay,
U. Ellwanger, S.~Moretti, and W.~Shepherd
are also gratefully acknowledged.
The author also thanks the 
late Professor G.~Soff for insightful 
discussions on general aspects of the 
true muonium bound system.

\appendix

%
%
\section{Coulomb--Gauge Propagator for Massive Vector Bosons}
\label{appa}

Even if the Coulomb-gauge propagator 
for massive vector bosons, given in Eqs.~\eqref{D00}
and~\eqref{Dij},
has been used in the literature before
[see Eq.~(6) of Ref.~\cite{VePa2004} and
Eqs.~(16) and (17) of Ref.~\cite{Je2011pra}],
separate notes on its derivation can clarify the role
of the ``Coulomb gauge'' for massive vector bosons.

Generally, in order to calculate the vector boson 
propagator in a specific gauge,
one can write the relation of the 
vector potential $A^\mu$ to the currents $J^\nu$ (in the gauge 
under investigation), and observe
that the operator mediating the relation is 
the propagator. However, one can also 
ask the question which terms could possibly 
be added to the propagator, initially obtained in a specific gauge,
without changing the fields.

Let us start with the massless case, i.e., the photon.
One should remember~\cite{Pa1993} that the 
most general form of a gauge transformation of the 
photon propagator is [in relativistic notation
with $k^\mu = (\omega, \vec k)$]
\begin{equation}
\label{gen_photon}
D_{\mu\nu}(k) = \frac{g_{\mu\nu}}{k^2} + \frac{1}{2 k^2} \left( f^\mu \, k^\nu +
f^\nu \, k^\mu \right) \,.
\end{equation}
The first term is the Feynman gauge result.
The added terms do not change the fields, 
because the term proportional to $f^\mu \, k^\nu$ vanishes
in view of current conservation, while the 
term proportional to $f^\nu \, k^\mu$ amounts to a gauge-transformation
of the four-vector potential $A^\mu$,
as a careful inspection shows.
The choice
\begin{equation}
f^0 = \frac{k^0}{\vec k^2} \,,
\qquad 
f^i = - \frac{k^i}{\vec k^2} \,,
\end{equation}
lead to the Coulomb gauge.
The fact that the fields remain unaffected holds
even for a ``non-covariant'' form of the 
$f^\mu$, i.e., for a form where the $f^\mu$ do not 
transform as components of a four-vector
under Lorentz transformations.

In order to generalize the result to a 
massive vector particle, it is necessary to 
observe that, in a more general sense, 
the ``Coulomb gauge'' for the calculation 
of bound states is defined to be the gauge in which 
the photon propagator component $D_{00}$ is exactly 
static, i.e., has no dependence on the photon 
frequency. Otherwise, one would incur 
additional corrections in the Breit Hamiltonian,
which is generated by the spatial components of the 
photon propagator. 

The generalization of Eq.~\eqref{gen_photon}
to the massive vector exchange reads as
[$g_{\mu\nu} = {\rm diag}(1,-1,-1,-1)$]
\begin{equation}
\label{gen_massive}
D_{\mu\nu}(k) = \frac{g_{\mu\nu}}{k^2 -m^2} + 
\frac{1}{2(k^2-m^2)} \left( f^\mu \, k^\nu +
f^\nu \, k^\mu \right) \,.
\end{equation}
Here, $m = m_X$ denotes the vector boson mass, 
and we start from the Feynman gauge result
$g_{\mu\nu}/(k^2 - m^2)$
[see Eqs.~(3.147) and~(3.149) in Ref.~\cite{ItZu1980},
with $\lambda = 1$, in the notation of Ref.~\cite{ItZu1980}].
Now, choosing 
\begin{equation}
f^0 = \frac{k^0}{\vec k^2 + m^2} \,,
\qquad
f^i = - \frac{k^i}{\vec k^2 + m^2} \,,
\end{equation}
one (almost) derives the result given in
Eqs.~\eqref{D00} and~\eqref{Dij}
[and previously used in Eq.~(6) of Ref.~\cite{VePa2004} and in
Eqs.~(16) and (17) of Ref.~\cite{Je2011pra}], with one caveat.
Namely, in the spatial components of the propagator (denoted 
by the Latin indices $ij$),
one replaces $k^2 = \omega^2 - \vec k^2 \to -\vec k^2$
in the order of approximation of interest here.
This is because the frequency dependence of the 
propagator denominator leads to higher-order
corrections, which, for the photon exchange,
are summarized in the Salpeter recoil correction~\cite{Sa1952}.

The transition
to the massive Coulomb gauge is well
known to be useful in bound-state theory
but is perhaps less familiar in the 
particle physics community; pertinent remarks are thus 
in order when it comes to the possible detection of a new
particle in low-energy experiments.

\end{document}